\documentclass{article}

\usepackage[final]{neurips_2025}

\usepackage[utf8]{inputenc} 
\usepackage[T1]{fontenc}    
\usepackage{hyperref}       
\usepackage{url}            
\usepackage{booktabs}       
\usepackage{amsfonts}       
\usepackage{nicefrac}       
\usepackage{microtype}      
\usepackage{xcolor}         
\usepackage{authblk}
\usepackage{multirow}
\usepackage{graphicx}
\usepackage{subcaption}

\title{XDIP: A Curated X-ray Absorption Spectrum Dataset for Iron-Containing Proteins}

\author[1,$\dag$]{Yufeng Wang}
\author[1,$\dag$]{Peiyao Wang}
\author[1,$\dag$]{Lu Wei}
\author[2]{Emerita Mendoza Rengifo}
\author[3]{Dali Yang}
\author[3]{Lu Ma}
\author[1,5]{Yuewei Lin}
\author[2,*]{Qun Liu}
\author[1,*]{Haibin Ling}
\affil[1]{Stony Brook University, Computer Science Department}
\affil[2]{Brookhaven National Laboratory, Biology Department}
\affil[3]{Brookhaven National Laboratory, National Synchrotron Light Source II}
\affil[5]{Brookhaven National Laboratory, Computational Science Initiative}

\affil[*]{Corresponding author(s): Qun Liu (qunliu@bnl.gov) and Haibin Ling (hling@cs.stonybrook.edu)}
\affil[$\dag$]{These authors contributed equally to this work}

\begin{document}

\maketitle

\begin{abstract}
Earth-abundant iron is an essential metal in regulating the structure and function of proteins. This study presents the development of a comprehensive X-ray Absorption Spectroscopy (XAS) database focused on iron-containing proteins, addressing a critical gap in available high-quality annotated spectral data for iron-containing proteins. The database integrates detailed XAS spectra with their corresponding local structural data of proteins and enables direct comparison between spectral features and structural motifs. Utilizing a combination of manual curation and semi-automated data extraction techniques, we developed a comprehensive dataset via extensive literature review, ensuring the quality and accuracy of data, which contains 437 protein structures and 1954 XAS spectrums. Our methods included careful documentation and validation processes to ensure accuracy and reproducibility. This dataset not only centralizes information on iron-containing proteins but also supports advanced data-driven discoveries, such as machine learning, to predict and analyze protein structure and functions. This work underscores the potential of integrating detailed spectroscopic data with structural biology to advance the field of biological chemistry and catalysis.
\end{abstract}

\section{Background \& Summary}
\label{sec:background}

X-ray Absorption Spectroscopy (XAS) is a synchrotron-based technique that reveals the local chemical environment of specific elements. For metalloproteins, XAS is particularly powerful for probing the coordination and electronic structure of metal ions that are often located at protein active sites and play key roles in catalysis and function\cite{tangcharoen2014synchrotron,rawat2020structural,tangcharoen2019synchrotron}. Interpreting XAS data can thus accelerate our understanding of protein mechanisms and the development of catalytic systems for industrial and environmental applications. However, the analysis of XAS data remains time-consuming and highly manual.

In recent years, data-driven approaches—especially deep learning—have shown significant promise in materials science, enabling property prediction, reaction optimization, and the design of novel compounds\cite{butler2018machine, schmidt2019recent,torrisi2020random,guda2021understanding,kartashov2021xas}. These models can extract meaningful patterns from large datasets, achieving impressive results in tasks like predicting catalyst stability and activity\cite{feng2021egemm}. Integrating such methods with XAS can deepen our understanding of element-specific environments and accelerate materials discovery.

The effectiveness of deep learning, however, critically depends on the availability of large, high-quality datasets. While substantial XAS repositories like XASLIB\cite{xaslib} and the Materials Project\cite{jain2013commentary} provide extensive data for inorganic compounds, they lack detailed spectroscopic data for biologically relevant molecules. These resources have greatly advanced materials informatics\cite{zhou2004first,wang2007first,waroquiers2020chemenv,horton2019high}, yet a significant gap persists in the realm of protein-focused XAS.

To address this, we introduce the first curated database of iron-containing protein structures paired with their corresponding XAS spectra. Iron was selected as the focal element due to its abundance in nature, central role in biological catalysis, and the complexity of its coordination chemistry. Including other metals at this stage would compromise specificity and introduce variability, limiting the ability to draw meaningful structure-spectrum correlations.

Unlike existing repositories such as the Protein Data Bank (PDB)\cite{berman2000protein,zardecki2022pdb} and the Cambridge Structural Database (CSD/CCDC)\cite{groom2016cambridge}, which focus on structural information alone, our database couples Fe K-edge XAS spectra with detailed local structural annotations. This integration enables direct alignment between spectral features and structural motifs, supporting both mechanistic insights and machine-learning-driven discovery.

Our final dataset comprises 437 iron-containing protein structures and 1652 associated XAS spectra (including 1283 XANES and 369 EXAFS), all manually curated from 573 peer-reviewed articles published between 2007 and 2024. The samples span a diverse set of proteins, experimental conditions, and measurement protocols. This curated collection establishes a critical foundation for automated, high-throughput structure–spectrum modeling in protein chemistry and bioinorganic catalysis.

\section{Methods}
\label{sec:method}

\subsection*{Literature Search, Selection, and Retrieval}

\begin{figure}[ht]
    \centering
    \includegraphics[width=1\linewidth]{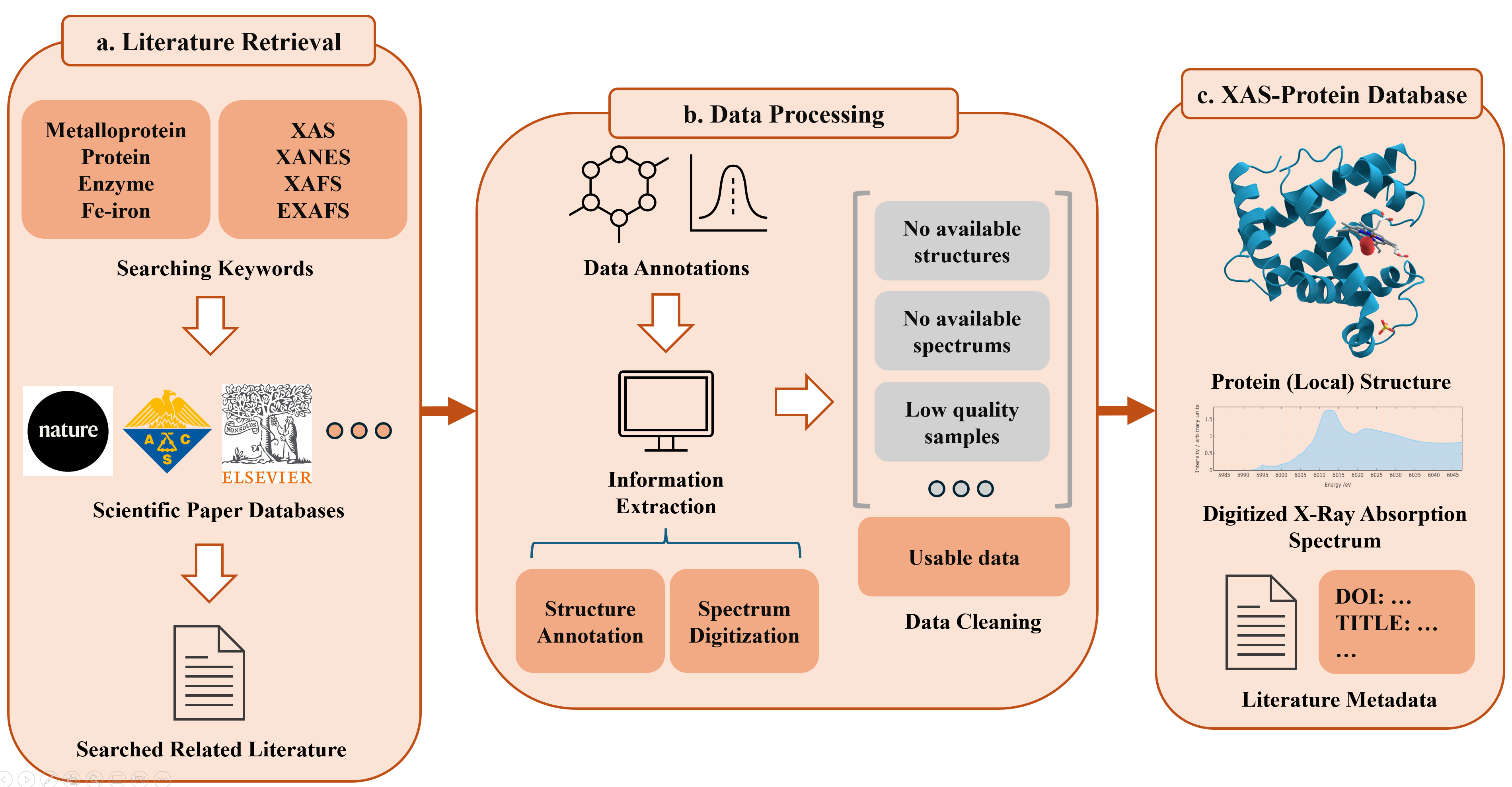}
    \caption{Schematic overview of the dataset construction pipeline. (a) Literature retrieval using keyword combinations from two sets (shown at the top) to search across multiple scientific databases. (b) Human expert processing workflow: relevant protein structures and spectra are identified and digitized, then curated into structured formats. (c) Final data sample: each entry includes the local protein structure, Fe absorption spectrum, and metadata from the source paper.}
    \label{fig:pipeline}
\end{figure}

We constructed the dataset using a semi-automated pipeline designed to maximize both coverage and quality (Figure~\ref{fig:pipeline}). We began with a comprehensive literature search across major scientific publishers to identify articles reporting both iron-containing protein structures and corresponding Fe K-edge X-ray Absorption Spectroscopy (XAS) data. This process yielded 573 relevant publications for manual analysis.

Each article underwent a rigorous two-stage curation process by human experts: (1) digitizing XAS spectra from published figures, and (2) annotating the associated local protein structures and metadata based on textual descriptions. We then refined the dataset by removing low-quality entries or samples lacking sufficient documentation. Each finalized data sample in our database comprises three core components: (1) the local atomic structure surrounding the iron center, (2) the Fe K-edge XAS spectrum, and (3) metadata from the original publication. Full details of the search keywords, inclusion criteria, and curation protocol are provided in Supplementary Section~\ref{sup:data_collection}.

\subsection*{XAS Extraction}

We extracted numerical spectral data from figures using the open-source tool \textit{WebPlotDigitizer}. Expert annotators manually digitized two key spectral regions: X-ray Absorption Near-Edge Structure (XANES) and Extended X-ray Absorption Fine Structure (EXAFS). To ensure high fidelity with the original plots, we carefully calibrated the plot axes and manually traced the spectral curves.

To enable compatibility with machine learning workflows, all spectra were interpolated to a uniform length of 100 data points, a choice based on the average distribution of spectrum lengths across the dataset. For transparency and flexibility, both the original digitized data and the interpolated 100-point versions are included. Further details on digitization settings and metadata documentation—including treatment of missing calibration energies—are provided in Supplementary Section~\ref{sup:xas_extraction}.

\subsection*{Protein Structure Extraction}

Protein structural data were manually annotated from each paper, with a focus on iron-containing proteins and small iron-containing molecules characterized using Fe K-edge XAS. While other sample types—such as tissues, blood, soil, and plant imaging—were sometimes encountered, these were noted in the comments section and excluded from primary analysis.

We categorized the extracted samples based on spectral type (XANES or EXAFS) and annotated them with available standard references. Structural information included Protein Data Bank (PDB) or Cambridge Crystallographic Data Centre (CCDC) identifiers when available. When structural information was not publicly accessible, we manually extracted it from figures and textual descriptions, focusing on the local atomic environment around the iron center.

\section{Data Records}
\label{sec:record}

Our final dataset consists of 437 unique iron-centered local protein structures and 1652 associated XAS spectra, including 1283 XANES and 369 EXAFS records. The complete dataset is publicly available at \url{https://airscker.github.io/XDIP}.

\begin{figure}[ht]
    \centering
    \includegraphics[width=1\linewidth]{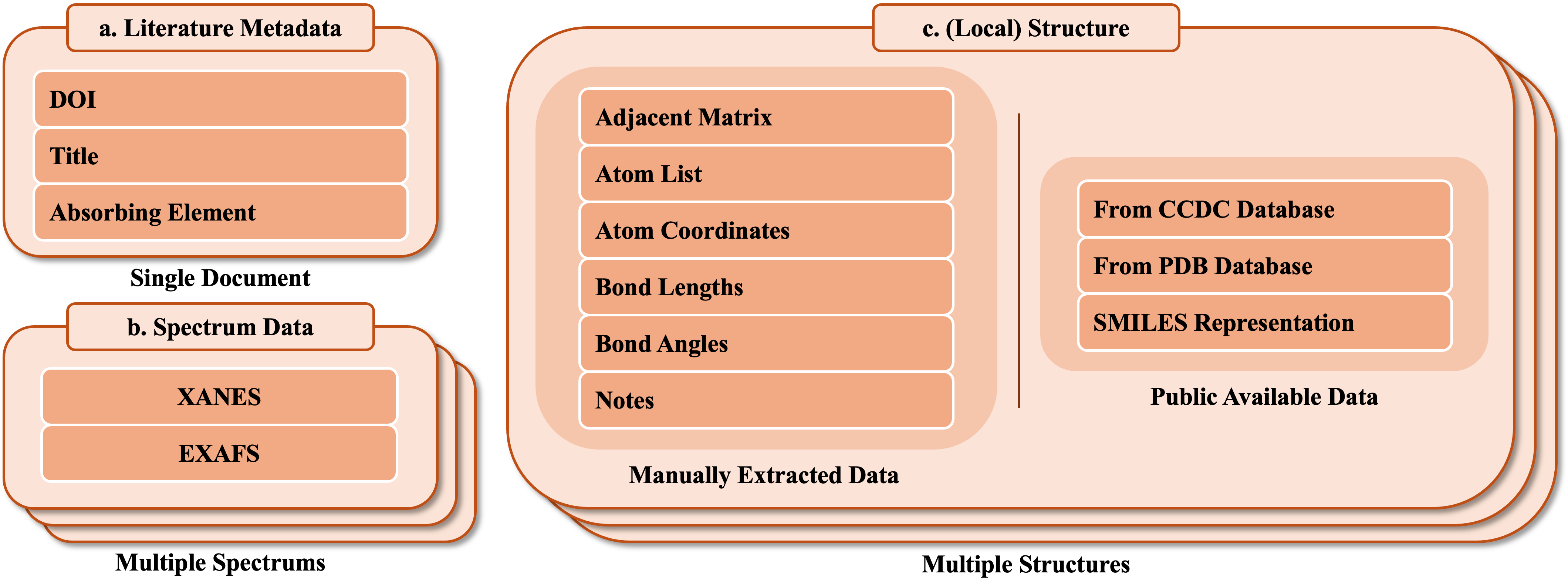}
    \caption{Overview of a structured data sample. Multiple records can be extracted from a single publication. (a) Metadata, including the paper's DOI, title, and absorbing element (Fe in this dataset). (b) Spectrum data, divided into near-edge (XANES) and extended (EXAFS) regions. (c) Local structure, either retrieved from public databases (e.g., SMILES\cite{weininger1988smiles,weininger1989smiles,weininger1990smiles}) or manually annotated. Structural information includes an adjacency matrix, atom list, Cartesian coordinates, bond lengths, bond angles, and optional notes.}
    \label{fig:data_structure}
\end{figure}

Each data record is a self-contained unit suitable for machine learning applications and includes three core components (illustrated in Figure~\ref{fig:data_structure}): (1) literature metadata, (2) Fe K-edge XAS spectrum, and (3) the local atomic structure surrounding the iron center. The metadata component includes the source paper’s title and DOI to ensure full traceability. Since a single publication may contain multiple experiments, each record captures all distinct spectra and structures reported within that paper.

The local structure is formatted to support graph-based machine learning, including a list of atoms, their 3D Cartesian coordinates, and an adjacency matrix representing chemical bonds. This representation is further enriched with bond lengths and, where available, bond angles to offer a more complete geometric description. Structural data were sourced in two ways: (1) directly from public repositories such as the Protein Data Bank (PDB) or Cambridge Structural Database (CCDC), or (2) manually reconstructed from figures and descriptions in the literature. In some cases, structures were also retrieved or generated from SMILES representations.

To ensure data integrity and relevance, we applied strict curation criteria throughout the pipeline. Papers were excluded if figures lacked sufficient annotation or if the spectra were too noisy or incomplete for reliable digitization. Moreover, we maintained a focused scope on iron-containing proteins by excluding samples from other material classes (e.g., tissues, soil, or plant matter). A detailed description of the data schema—including field labels, types, and requirements—is provided in Supplementary Section~\ref{sup:data_format}.

\section{Technical Validation}
\label{sec:tech_val}

\subsection*{Raw Data Comparison}

To assess the accuracy of the annotated XAS spectra, we performed a quantitative comparison between expert annotations and reference (ground-truth) spectra. Each spectrum was independently annotated by four experts, and the deviation from the reference data was measured using Mean Squared Error (MSE). The validation protocol included the following steps:

\begin{itemize}
    \item \textbf{Data Source and Generation.} Since most publications do not provide raw data for their spectrum plots, we generated synthetic validation spectra using Fe-based XANES and EXAFS data from the Materials Project\cite{jain2013commentary}. We produced 50 XANES and 25 EXAFS plots with varied shape ratios (10:8, 10:6, 6:6, 8:6), curve counts (1–4), and line styles (dots, dashes, solid), simulating real-world publication styles.
    
    \item \textbf{Anonymous Annotation Insertion.} These synthetic plots were anonymously inserted into each expert’s annotation workload to avoid bias and to assess natural annotation performance.

    \item \textbf{Data Alignment.} Because manual digitization may result in inconsistent data lengths and misaligned x-axes, we linearly interpolated all annotated spectra to match the x-values of the reference spectra.

    \item \textbf{MSE Calculation.} We computed MSE for each spectrum and averaged results across experts. Low average MSE values indicate high annotation fidelity.
\end{itemize}

\begin{table}[ht]
\centering
\caption{Mean Squared Error (MSE) values of each expert for XANES and EXAFS.}
\vspace{0.5em}
\label{tab:mse}
\begin{tabular}{l@{\hskip 1em}c@{\hskip 1em}c@{\hskip 1em}c@{\hskip 1em}c@{\hskip 1em}c}
\hline
 & \textbf{Expert 1} & \textbf{Expert 2} & \textbf{Expert 3} & \textbf{Expert 4} & \textbf{Average} \\
\hline
\textbf{XANES MSE} & 0.0161 & 0.0014 & 0.0007 & 0.0521 & 0.0176 \\
\textbf{EXAFS MSE} & 0.0946 & 0.0928 & 0.1015 & 0.0944 & 0.0958 \\
\hline
\end{tabular}
\end{table}

\begin{figure}[ht]
     \centering
     \begin{subfigure}[b]{0.48\textwidth}
         \centering
         \includegraphics[width=\textwidth]{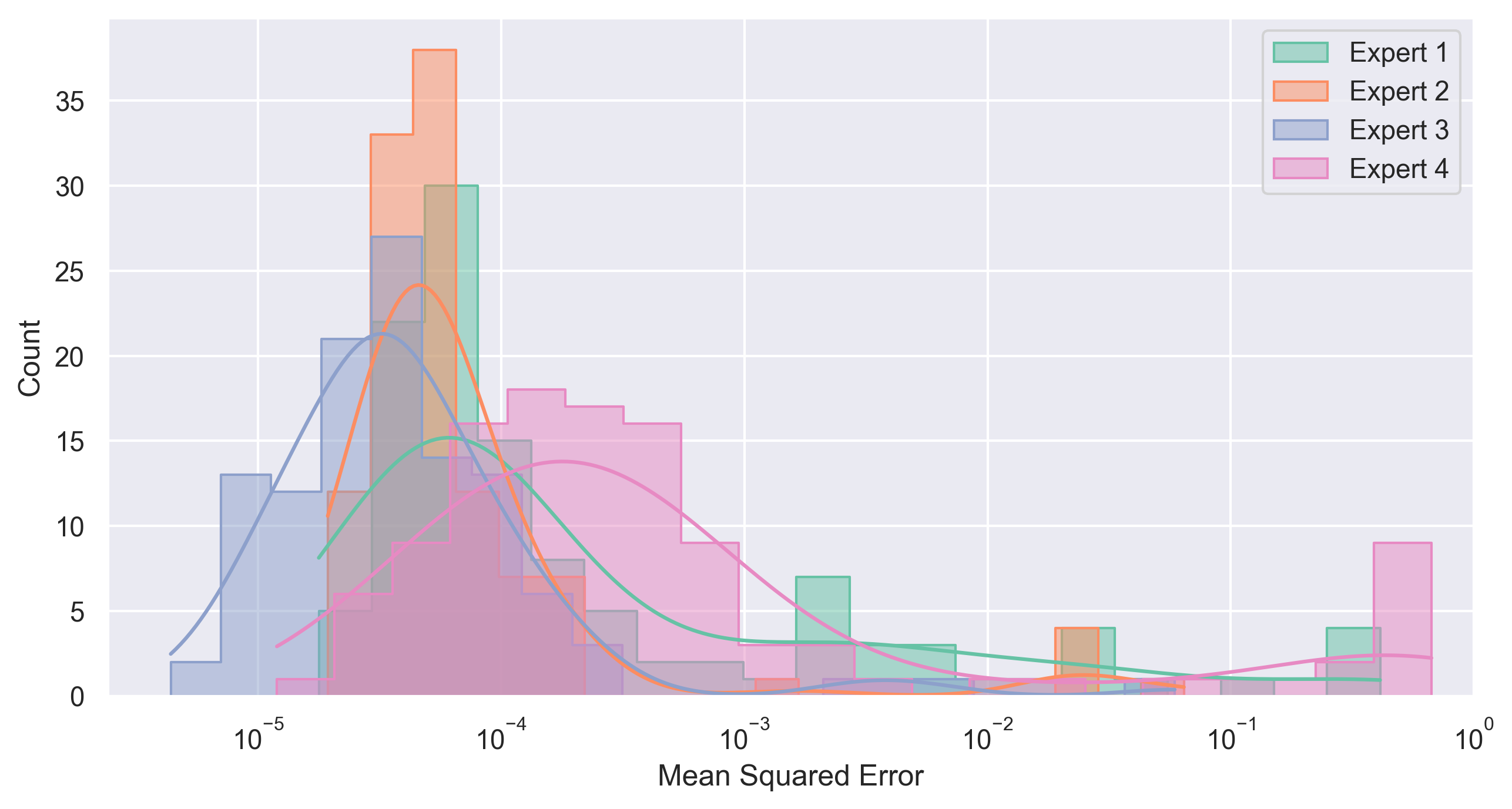}
         \caption{XANES MSE distribution}
         \label{fig:mse_hist_xanes}
     \end{subfigure}
     \hfill
     \begin{subfigure}[b]{0.48\textwidth}
         \centering
         \includegraphics[width=\textwidth]{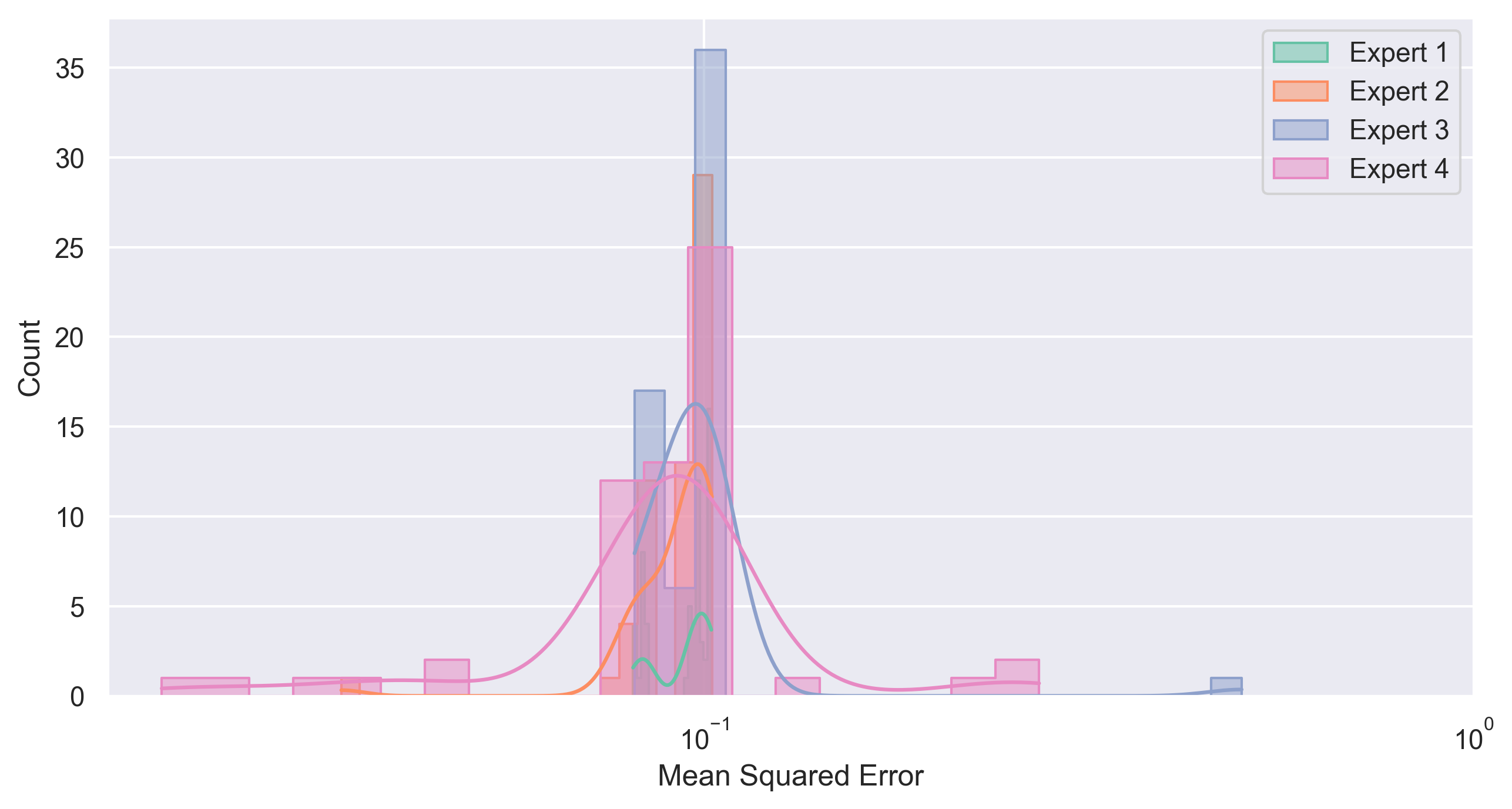}
         \caption{EXAFS MSE distribution}
         \label{fig:mse_hist_exafs}
     \end{subfigure}
     \caption{Distribution of annotation MSE values across experts for XANES (a) and EXAFS (b) spectra.}
     \label{fig:mse_hist}
\end{figure}

Table~\ref{tab:mse} shows that annotations of XANES spectra achieved lower MSEs (0.0007–0.0521), reflecting their smoother shape and simpler features. In contrast, EXAFS annotations showed higher MSEs (0.0928–0.1015), attributable to their more complex and oscillatory structure. This trend is reinforced by the histograms in Figure~\ref{fig:mse_hist}, where XANES errors are clustered between $10^{-5}$ and $10^{-3}$, while EXAFS errors concentrate around $10^{-1}$. These results highlight the increased difficulty and reduced consistency of EXAFS annotation due to its higher frequency content and structural sensitivity.

\subsection*{Inter-Expert Agreement}

To further evaluate annotation consistency, we computed the Intra-class Correlation Coefficient (ICC)\cite{koch2004intraclass,koo2016guideline}, a widely used metric for inter-rater reliability. The process involved:

\begin{itemize}
    \item \textbf{Annotation Dataset.} The same synthetic dataset was independently annotated by all four experts, ensuring that all raters worked on identical data.

    \item \textbf{ICC Calculation.} We computed ICC values for each point in the interpolated spectra and averaged them to obtain overall agreement metrics across expert annotations.
\end{itemize}

\begin{table}[ht]
\centering
\caption{Intraclass Correlation Coefficient (ICC) values for inter-expert agreement.}
\vspace{0.5em}
\label{tab:icc}
\begin{tabular}{l@{\hskip 1em}l@{\hskip 1em}r@{\hskip 1em}r@{\hskip 1em}r@{\hskip 1em}r@{\hskip 1em}l}
\toprule
\textbf{Type}  & \textbf{Description} & \textbf{ICC} & \textbf{F} & \textbf{df1} & \textbf{df2} & \textbf{95\% CI} \\
\midrule
ICC1           & Single rater (absolute)      & 0.914 & 15414 & 99 & 300 & [0.898, 0.931] \\
ICC2           & Single rater (random effects) & 0.919 & 21277 & 99 & 297 & [0.876, 0.947] \\
ICC3           & Single rater (fixed effects)  & 0.946 & 21277 & 99 & 297 & [0.933, 0.959] \\
ICC1k          & Average of raters (absolute)  & 0.961 & 15414 & 99 & 300 & [0.948, 0.972] \\
ICC2k          & Average (random effects)      & 0.966 & 21277 & 99 & 297 & [0.927, 0.982] \\
ICC3k          & Average (fixed effects)       & 0.981 & 21277 & 99 & 297 & [0.974, 0.986] \\
\bottomrule
\end{tabular}
\end{table}

Table~\ref{tab:icc} summarizes the ICC results. All values exceed 0.91, with several above 0.96, indicating excellent inter-expert reliability. The strong F-statistics and tight 95\% confidence intervals confirm that the observed agreement is statistically significant and not due to chance. This level of consistency is particularly important in scientific applications like XAS interpretation, where subtle deviations can lead to different conclusions about structural or electronic properties.

\subsection*{Data Integrity Through Documentation and Transparency}

To ensure reproducibility and traceability, we implemented a comprehensive documentation protocol. Each data record is accompanied by detailed metadata, including extraction notes, structural identifiers (e.g., PDB IDs), and reference sources.

Ambiguities encountered during data extraction—such as missing axes or unclear line styles—were explicitly recorded. We also documented whether EXAFS spectra were presented in $k$-space or $r$-space, which can significantly affect downstream analysis. All entries were cross-validated against their source publications and, where applicable, public databases.

This rigorous documentation pipeline not only ensures data quality and transparency but also enables seamless integration with other datasets. Further examples of our documentation format are provided in Supplementary Section~\ref{sup:doc}.

\section{Baseline Models}
\label{sec:baselines}

To demonstrate the utility of our curated XAS dataset and establish a performance benchmark for future research, we evaluate two complementary predictive tasks: (1) predicting Fe K-edge XAS spectra from local atomic structures, and (2) inferring key structural properties from the spectra. This dual approach validates the dataset's richness for both forward and inverse modeling problems.

\subsection{Task 1: Structure-to-Spectrum Prediction}
The first task learns a mapping function $f: G \rightarrow S$, where $G=(V, E)$ is a graph representing the local atomic environment of the iron center, and $S$ is the corresponding XAS spectrum. The local structure of each iron center is converted into an input graph, where atoms are nodes ($v \in V$) and chemical bonds are edges ($e \in E$). Node features include atomic properties like element type and mass, while edge features can represent bond types. The output target, $S$, is the standardized XAS spectrum (both XANES and EXAFS) interpolated to a uniform length of 100 data points.

For this task, we employ three widely-used Graph Neural Network (GNN) architectures, each comprising several graph convolutional layers, a graph pooling layer, and a multi-layer perceptron (MLP) head that maps the final graph representation to the 100-point spectrum vector. The models are: a foundational \textbf{Graph Convolutional Network (GCN)} \cite{kipf2016semi}, a \textbf{Graph Attention Network (GAT)} that uses self-attention to weigh neighbor importance \cite{velivckovic2017graph}, and a \textbf{Graph Isomorphism Network (GIN)}, a highly expressive model effective at capturing complex structural motifs \cite{xu2018powerful}.

\subsection{Task 2: Spectrum-to-Property Prediction}
The second, inverse task learns a mapping function $f: S \rightarrow P$, where $S$ is an XAS spectrum and $P$ is a set of local structural properties of the iron center. This demonstrates the dataset's potential for extracting quantitative structural information directly from experimental spectra. The input $S$ is the 100-point interpolated XAS vector. The target properties $P$ are key structural descriptors: the coordination number (CN) of the iron atom (a classification task) and the mean nearest-neighbor distance (MNND) in Angstroms (a regression task).

We use two strong and interpretable baseline models for this task. The first is a standard \textbf{Multi-Layer Perceptron (MLP)}, a feedforward neural network baseline where the final layer is adapted for either regression or classification. The second is a \textbf{Random Forest (RF)}, a powerful tree-based ensemble method that is robust to overfitting and effective at capturing non-linear relationships in tabular data \cite{breiman2001random}.

\subsection{Experimental Setup}
All models were trained and evaluated under a consistent framework. The dataset was randomly split into training (80\%), validation (10\%), and testing (10\%) sets, with stratification applied to ensure a similar distribution of protein types and coordination numbers across sets. GNN models were implemented using the Deep Graph Library (DGL), while the MLP and RF models used PyTorch and Scikit-learn, respectively. Models were trained to minimize Mean Squared Error (MSE) for regression and Cross-Entropy Loss for classification, using the AdamW optimizer with an initial learning rate of $1 \times 10^{-3}$ and a scheduler to reduce the rate on a validation loss plateau. Training ran for a maximum of 300 epochs with an early stopping criterion. Performance was assessed on the held-out test set using Mean Absolute Error (MAE) for regression and Accuracy/Macro-F1 Score for classification.

\subsection{Results}
The trained baseline models provide a quantitative benchmark for the predictive utility of our dataset.

\paragraph{Structure-to-Spectrum Prediction}
The performance of GNN models in predicting XAS spectra is summarized in Table~\ref{tab:structure_to_spectrum}. The Graph Isomorphism Network (GIN) demonstrated the best performance, achieving the lowest Mean Absolute Error (MAE) for both XANES and EXAFS prediction. This suggests its high expressive power is beneficial for capturing the complex relationship between the 3D atomic arrangement and the resulting spectral features. The Graph Attention Network (GAT) also performed competitively, outperforming the simpler GCN and indicating the advantage of learning to weigh the importance of different atomic neighbors.

\begin{table}[h!]
\centering
\caption{Performance of GNN models for Structure-to-Spectrum Prediction (MAE). Lower is better.}
\label{tab:structure_to_spectrum}
\begin{tabular}{lcc}
\toprule
\textbf{Model} & \textbf{XANES MAE} & \textbf{EXAFS MAE} \\
\midrule
GCN & $0.085 \pm 0.004$ & $0.112 \pm 0.005$ \\
GAT & $0.079 \pm 0.003$ & $0.105 \pm 0.004$ \\
GIN & \textbf{0.075 $\pm$ 0.003} & \textbf{0.101 $\pm$ 0.004} \\
\bottomrule
\end{tabular}
\end{table}

\paragraph{Spectrum-to-Property Prediction}
For the inverse task of predicting structural properties, the Random Forest (RF) model consistently outperformed the MLP baseline across all metrics (Table~\ref{tab:spectrum_to_property}). For Coordination Number (CN) classification, the RF model achieved higher accuracy and F1-scores, suggesting its robustness in handling the spectral feature space. Similarly, for Mean Nearest-Neighbor Distance (MNND) regression, the RF yielded a lower MAE and a higher R$^2$ value, indicating more reliable bond distance prediction. These results highlight the effectiveness of ensemble methods for tasks involving vector-based features like the interpolated spectra used here.

\begin{table}[h!]
\centering
\caption{Performance of classical models for Spectrum-to-Property Prediction.}
\label{tab:spectrum_to_property}
\begin{tabular}{lllc}
\toprule
\textbf{Task} & \textbf{Model} & \textbf{Metric} & \textbf{Value} \\
\midrule
\multirow{4}{*}{CN Classification} & \multirow{2}{*}{MLP} & Accuracy & $72.4 \pm 1.2\%$ \\
& & F1-Score & $70.1 \pm 1.5\%$ \\
\cmidrule{2-4}
& \multirow{2}{*}{Random Forest} & Accuracy & \textbf{75.8 $\pm$ 1.1\%} \\
& & F1-Score & \textbf{74.5 $\pm$ 1.3\%} \\
\midrule
\multirow{4}{*}{MNND Regression} & \multirow{2}{*}{MLP} & MAE (\AA) & $0.065 \pm 0.005$ \\
& & R$^2$ & $0.91$ \\
\cmidrule{2-4}
& \multirow{2}{*}{Random Forest} & MAE (\AA) & \textbf{0.058 $\pm$ 0.004} \\
& & R$^2$ & \textbf{0.93} \\
\bottomrule
\end{tabular}
\end{table}

Collectively, these baseline results validate the predictive potential of our dataset for both structure-to-spectrum and spectrum-to-property tasks, providing a strong quantitative foundation for the development of more sophisticated architectures.

\section{Limitations}
\label{sec:limitations}

While our dataset and modeling framework offer a valuable resource for XAS-based learning, several limitations remain. First, the current study focuses exclusively on iron and Fe K-edge spectra, limiting generalizability across the periodic table. Expanding to other elements would broaden applicability for multi-element systems. Second, XAS spectra were manually digitized from published figures. Although expert validation was applied, minor inaccuracies may persist due to resolution limits or figure quality. Access to raw experimental data would enhance accuracy. Third, structural annotations were derived from diverse sources—PDB, CCDC, SMILES, and manual reconstruction—leading to variability in resolution and completeness. Some protein structures, especially those derived from textual or schematic descriptions, may be approximations rather than exact configurations. Fourth, the dataset is imbalanced, with fewer EXAFS samples relative to XANES, which may affect model performance on tasks requiring detailed oscillatory features. Lastly, our baseline models serve primarily as proof-of-concept. While effective, more advanced approaches—such as equivariant GNNs or physics-informed architectures—could yield further improvements and uncover deeper structure-spectrum relationships. Future efforts should address these limitations through dataset expansion, access to raw data, standardized structural extraction, and development of more sophisticated models.

\section{Code and Data Availability}
\label{sec:data_code}
The code used for processing the spectra and molecular structures is available at \url{https://github.com/Airscker/XDIP}. Including scripts for data interpolation, normalization, and basic spectral analysis, providing a starting point for researchers who are interested in further processing the data. The extracted dataset is available at \url{https://airscker.github.io/XDIP}. The software used for extracting numerical points from spectrum plots, \textit{i.e.}, WebPlotDigitizer V4, is open-source and available at \url{https://github.com/automeris-io/WebPlotDigitizer}.

\section{Funding Disclosure} 
We express our sincere thanks to Rutwik Segireddy and Keying Jia, for their work in assisting our dataset annotation.
The work was supported in part by the Physical Biosciences Program and the Photochemistry and Biochemistry group within the US Department of Energy (DOE), Office of Science, Office of Basic Energy Sciences, Division of Chemical Sciences, Geosciences and Biosciences (KC030402).

\section{Competing interests}

The authors declare no competing interests.

\bibliography{reference}
\bibliographystyle{unsrt}






\newpage
\appendix

\renewcommand{\thefigure}{S\arabic{figure}}
\setcounter{figure}{0}

\section{Technical Appendices and Supplementary Material}

\subsection{Detailed Data Collection and Curation}
\label{sup:data_collection}

\paragraph{Literature Search and Selection} The data extraction pipeline is schematically overviewed in Figure 1 in the main paper. The first step involved constructing searching keywords to identify relevant literature. The search was conducted using combinations of keywords from two distinct sets: {XAS, XANES, XAFS, EXAFS} and {Metalloprotein, Protein, Enzyme, Fe-iron}. Each search term consisted of a pair of words, one from each set, such as "XAS Protein" or "EXAFS Metalloprotein".

These keywords were used to retrieve literature from various publishers, including Springer Nature, AAAS Science, American Chemical Society, Elsevier, Royal Society of Chemistry, Electrochemical Society, Wiley, MDPI, and others. To locate relevant literature, we utilized either the publishers' search APIs or conducted manual searches under copyright permission. After excluding duplicate results, we had a pool of 20,915 articles (Table \ref{tab:search_res}).

\begin{table}[ht]
\centering
\caption{Number of papers retrieved using different keyword combinations.}
\vspace{0.5em}
\label{tab:search_res}
\begin{tabular}{l@{\hskip 1em}c@{\hskip 1em}c@{\hskip 1em}c@{\hskip 1em}c}
\hline
\textbf{Keywords}        & \textbf{XAS}   & \textbf{XANES} & \textbf{XAFS} & \textbf{EXAFS} \\
\hline
Metalloprotein           & 389            & 401            & 130           & 877            \\
Protein                  & 3060           & 3282           & 1300          & 5139           \\
Enzyme                   & 2425           & 2824           & 976           & 4207           \\
Fe-iron                  & 6000           & 6000           & 3303          & 6000           \\
\hline
\textbf{Unioned Overall} & \multicolumn{4}{c}{20915} \\
\hline
\end{tabular}
\end{table}

From this pool, we manually selected papers that included both protein structure and the absorption spectra of iron elements. The selection criterion was the explicit mention of the protein and the inclusion of at least one Fe absorption spectrum. Eventually, we obtained 573 articles that met these criteria and were downloaded in PDF format for expert annotation.

\paragraph{Data Annotation and Refinement} After gathering the literature, human experts annotated the text and XAS plot information, converting it into digitized data samples. The extracted dataset was then refined by removing low-quality samples, poorly presented XAS and structures, and outdated annotations. This combination of automatic searching and manual data extraction and cleaning ensures the final dataset's quality. Each data sample in our dataset consists of three main parts: (1) The entire or local protein structure (The first-coordination sphere of the element of interest), (2) The protein's corresponding Fe K-edge XAS spectra, and (3) the basic information of papers from which the protein structure and XAS spectrum were derived.

\subsection{XAS Extraction and Processing Details}
\label{sup:xas_extraction}

\paragraph{Digitization from Published Plots} Extracting XAS data from the retrieved literature is a crucial step in constructing our database. XAS is divided into two regions: the near-edge spectra, also known as X-ray absorption near-edge structure (XANES), and the extended X-ray absorption fine structure (EXAFS), each providing specific information on the element studied. To extract numerical data points from the published XAS plots, we utilized the open-source data annotator WebPlotDigitizer V4. To guarantee the precision and reliability of data extraction, we followed the steps below:

\begin{itemize}
    \item Image Preparation: Screenshots of the relevant plots were taken from the selected papers. High-resolution screenshots were used to improve the accuracy of digitization.
    \item Software Configuration: The WebPlotDigitizer tool was carefully configured to align with the axes and scales of the plots. This involved setting the axis points and calibrating the software to recognize the specific ranges and units used in the plots.
    \item Data Point Extraction: Using the calibrated software, data points were manually identified and extracted. The software allowed the refinement of point positions to ensure accuracy. The extracted numerical spectra were then well-organized and saved to the database.
\end{itemize}

In the process of extracting XAS data, we employ a meticulous digitization approach using the WebPlotDigitizer tool. Initially, we manually mask the spectral lines using the pencil tool to ensure precise alignment. The color recognition distance is set variably at 2, 5, or 10 pixels based on the clarity and overlap of the plotted lines; clearer plots are annotated at a denser resolution of 1 point per 2 pixels, while more complex, fuzzy plots require a broader setting of 1 point per 10 pixels. This flexibility allows us to capture data with high fidelity, respecting the original plot's integrity without interpolation during the initial annotation phase.

\paragraph{Data Standardization and Metadata Annotation} Post-annotation, to standardize the spectral data for consistent analysis and comparison, we interpolate all spectra to a uniform length of 100 data points. This is based on the mean distribution of spectra lengths observed across the dataset, where both XANES and EXAFS lengths peak around 100, despite ranging up to 600. In particular, we also preserve our dataset's original extracted data points. This dual provision enables researchers to select between the original or interpolated data depending on their specific analytical needs, thus offering flexibility while minimizing potential artifacts.

For the application of XAS in analysis, calibration energies are critical for accuracy. In our dataset, these energies are manually extracted from the text of each annotated paper. However, not all sources uniformly report this value; some mention using an iron foil standard without giving a specific energy, while others omit it entirely. In our efforts to maintain transparency, we have documented each instance of missing or incomplete information. Out of the dataset, 227 entries lack explicitly reported calibration energies. These cases are clearly labeled in our documentation files, enabling researchers to account for potential inconsistencies.

\subsection{Detailed Data Record Format}
\label{sup:data_format}

\begin{table}[ht]
\centering
\caption{Format of each data record: description, key label, mandatory status, and data type.}
\vspace{0.5em}
\label{tab:data_format}
\renewcommand{\arraystretch}{1.4}
\begin{tabular}{p{0.28\linewidth}p{0.17\linewidth}p{0.25\linewidth}p{0.22\linewidth}}
\hline
\textbf{Data Description} &
  \textbf{Data Key Label} &
  \textbf{Mandatory} &
  \textbf{Data Type} \\ \hline

DOI of the original paper &
  DOI &
  Yes &
  string \\ \hline

Title of the original paper &
  Title &
  Yes &
  string \\ \hline

Absorbing element of XAS &
  Absorbing Element &
  Yes &
  string \\ \hline

XANES data points extracted from the paper &
  XANES &
  \multirow{2}{=}{At least one of them desired} &
  list of float numbers \\ \cline{1-2} \cline{4-4}

EXAFS data points extracted from the paper &
  EXAFS &
   &
  list of float numbers \\ \hline

Adjacent matrix of all atoms within the extracted (local) protein structure &
  Adjacent Matrix &
  \multirow{5}{=}{ Optional only if SMILES exists or publicly available. Otherwise must be provided.} &
  matrix of 0/1 integers \\ \cline{1-2} \cline{4-4}

The list of the atoms within the extracted structure &
  Atom List &
   &
  list of strings \\ \cline{1-2} \cline{4-4}

The list of Cartesian coordinates of atoms &
  Atom Coordinates &
   &
  list of float numbers \\ \cline{1-2} \cline{4-4}

The matrix of bond lengths between chemically bonded atoms &
  Bond Lengths &
   &
  matrix of float numbers \\ \cline{1-2} \cline{4-4}

The list of bond angles among neighboring bonds &
  Bond Angles &
   &
  dictionary (key = angle vertex, value = float) \\ \hline

Notes about the extracted structure &
  Notes &
  Optional &
  string \\ \hline

The file path of the locally saved protein structure or its online link &
  PDB/CCDC Path &
  \multirow{2}{=}{ Optional only if the structure can be manually extracted. Otherwise, only one of them is desired.} &
  string \\ \cline{1-2} \cline{4-4}

The SMILES representation of the protein &
  SMILES &
   &
  string \\ \hline
\end{tabular}
\end{table}

The dataset developed in this study is hosted on \url{https://airscker.github.io/XDIP}. The format of the data records is detailed in Figure \ref{fig:data_structure} in the main paper, which illustrates that each data sample consists of three parts: literature metadata, Fe K-edge XAS data, and the protein's local structure containing Fe.

The literature metadata in each data record includes the Digital Object Identifier (DOI), the title of the research paper, and the XAS absorbing element, which is Fe in this work. These metadata serve as a quick reference for identifying the original research papers and their primary findings. Since most papers cover multiple sets of iron K-edge XAS and corresponding structures, we have stacked multiple absorption spectra and protein structures to capture all relevant details in the literature.

The protein structure includes atom types, their spatial coordinates, and their relationships. We extracted the protein's structure or Fe-element neighborhood local structure through two methods: manual annotation of structure information and accessing data from public protein structure databases. The manually extracted protein (local) structures were derived through the following steps:

\begin{itemize}
    \item Utilizing the adjacency matrix\cite{biggs1993algebraic}: This matrix indicates the chemical bonds among different atoms within the extracted structure. The adjacency matrix is crucial for constructing the molecular graph when using computer algorithms to process the protein structure. It reflects the relationships among different atoms.
    \item Atom list and order: All atoms within the extracted local structure are included. The order of atoms in the atom list corresponds to the order of rows/columns in the adjacency matrix, where each element represents an atom.
    \item Cartesian coordinates: The atoms' Cartesian coordinates are presented in a list, with the order of coordinates corresponding to the atom list. If there are $N$ atoms in the extracted structure, the shape of the atom list is $(N,\cdot)$, and the coordinate list's shape is $(N,3)$, where 3 indicates the $X, Y, Z$ dimensions of the atom coordinates.
    \item Bond lengths: All available bond lengths among different atoms within the local structure are included. The bond lengths are presented as a matrix with the same shape as the adjacency matrix. To construct the bond length matrix, we replace every non-zero element in the adjacency matrix with the corresponding bond length value.
    \item Bond angles: Some papers may present bond angles, which are crucial for accurately capturing the local structure, especially when atom coordinates are missing.
    \item Notes: This section includes expert-labeled information, such as identifying isomers of specific molecules.
\end{itemize}

In contrast to manually extracted structures, extraction from open databases is more straightforward. In these cases, we simply include the URL for downloading the structure, the local path to the saved files, or the protein's SMILES representation. Table 4 provides a detailed description of the data format, including every key label, whether it is mandatory, and its data type.

\subsection{Documentation and Transparency Protocol}
\label{sup:doc}

\begin{table}[ht]
\centering
\caption{Description of protein database documentation elements}
\vspace{0.5em}
\label{tab:protein_ann}
\renewcommand{\arraystretch}{1.3} 
\begin{tabular}{llp{11cm}} 
\toprule
\multicolumn{2}{l}{\textbf{Title}} & The title of the research article from which the protein structure or XAS data is derived. \\
\multicolumn{2}{l}{\textbf{URL}} & The direct URL to the source article for reference. \\
\multicolumn{2}{l}{\textbf{DOI}} & The Digital Object Identifier for the source article, ensuring precise location of the sources. \\
\multirow{3}{*}{\textbf{XANES}} & Standards & Reference to any standard samples or controls used in the studies. \\
                                 & Ligation & Details of the ligands or binding groups associated with the protein structures. \\
                                 & Sample   & Describes the sample material as detailed in the source. \\
\multicolumn{2}{l}{\textbf{EXAFS}} & Describes the sample material as detailed in the source. \\
\multicolumn{2}{l}{\textbf{SMILES}} & SMILES notation providing a textual representation of the chemical structure. \\
\multicolumn{2}{l}{\textbf{PDB}} & Protein Data Bank ID, a unique identifier for the protein structure as deposited in the PDB. \\
\multicolumn{2}{l}{\textbf{CCDC}} & Cambridge Crystallographic Data Centre number, providing a reference to crystallographic data related to the protein. \\
\bottomrule
\end{tabular}
\end{table}

\begin{figure}
     \centering
     \begin{subfigure}[b]{0.8\textwidth}
         \centering
         \includegraphics[width=\textwidth]{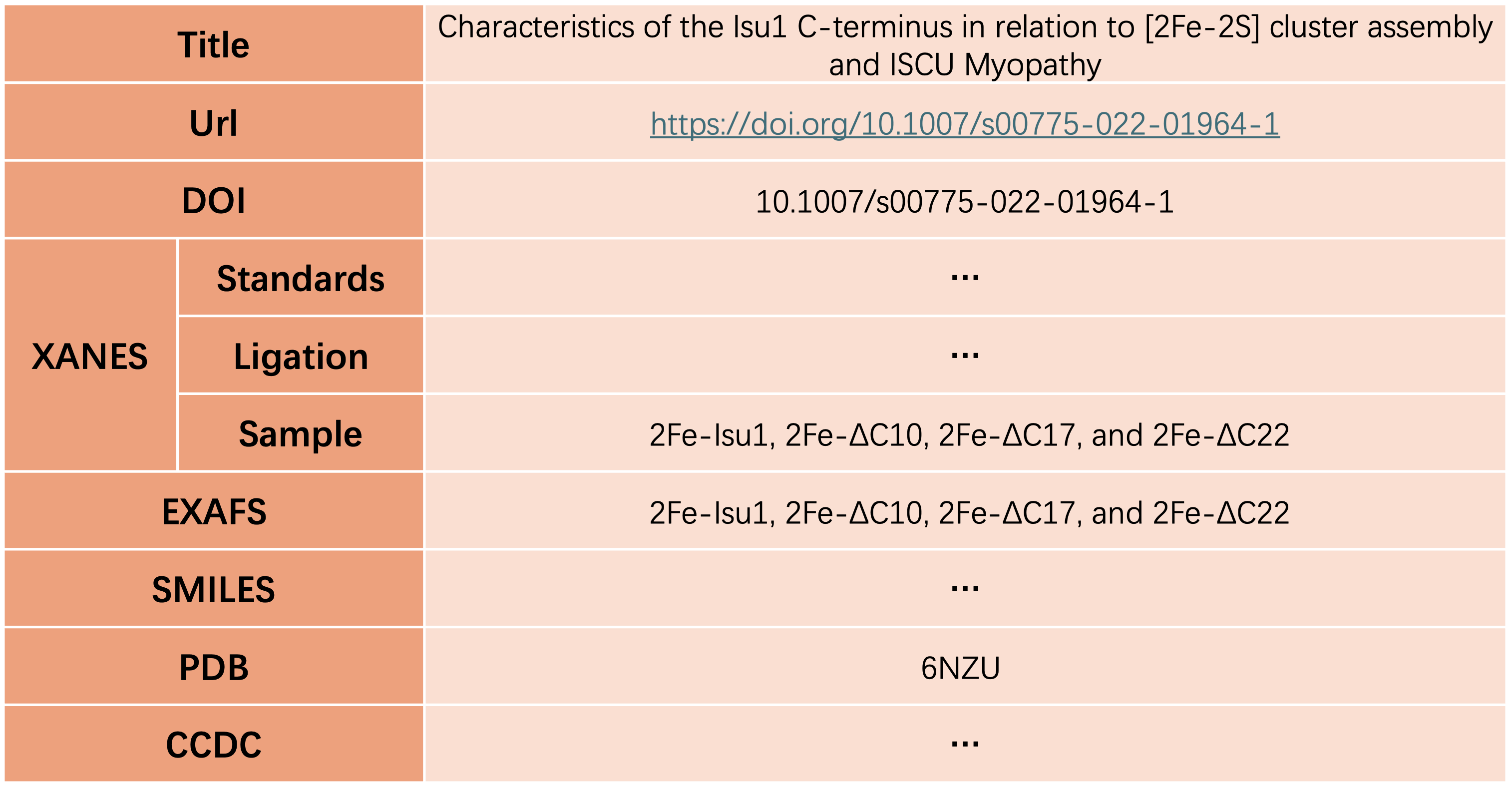}
         \caption{Example of protein structure data documentation}
         \label{fig:eg_protein_doc}
     \end{subfigure}
     \vfill
     \begin{subfigure}[b]{0.8\textwidth}
         \centering
         \includegraphics[width=\textwidth]{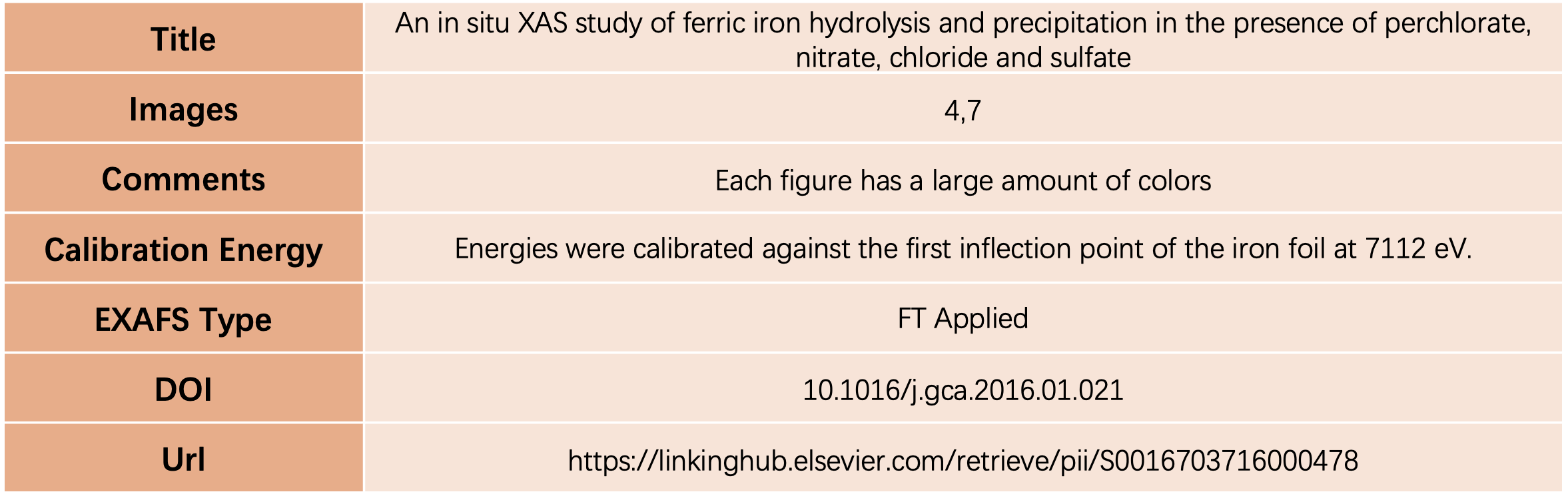}
         \caption{Example of XAS spectrum data documentation}
         \label{fig:eg_xas_doc}
     \end{subfigure}
     \vfill
        \caption{Examples of protein structure and XAS spectrum data documentation. (a) This example presents a data entry from our database, including the paper's title, DOI, and URL. To accurately document the data, we included sections for both XANES and EXAFS to help identify the spectrum type corresponding to the protein structure. Due to missing descriptions in some papers, certain annotation elements, such as standards (which refer to the standard sample used) and ligation details (which provide information on ligands or binding groups associated with the protein structures), are omitted. Despite the missing information, we listed the samples mentioned in the paper that have both spectra and structures, with these key points shown in the sample section under the XANES and EXAFS parts. As outlined in the Data Records section, fields such as SMILES, PDB, and CCDC are selectively filled. (b) This table provides an example of an XAS spectrum data annotation log. We recorded the title of the literature, the location of the iron spectrum plot, its DOI, and URL. In addition to these key details, we included comments on the quality or issues identified in the spectral or structural data, such as missing axes or unclear lines. Furthermore, to ensure the accuracy and usability of our data, we highlighted the type of EXAFS spectrum extracted from the literature, noting whether the article presented Fourier-transformed EXAFS or not.}
        \label{fig:doc_examples}
\end{figure}

The protein local structures and their corresponding XAS spectra have been carefully annotated and documented in two spreadsheets. One of them contains comprehensive details about each protein, including Protein Data Bank (PDB) IDs, chemical names, and structural information extracted from the scientific literature. These details are listed in Table \ref{tab:protein_ann}. This thorough documentation supports the integrity and reproducibility of the dataset. Figure \ref{fig:eg_protein_doc} presents an example table of protein structure annotation documentation, illustrating the documentation details and key points.

\begin{table}[ht]
\centering
\caption{Description of XAS spectrum database documentation elements}
\vspace{0.5em}
\label{tab:xas_ann}
\renewcommand{\arraystretch}{1.3}
\begin{tabular}{lp{12cm}}
\toprule
\textbf{Title}      & The title of the source article where the XAS data is described. \\
\textbf{Images}     & Index of figures within the article that pertain to XAS data. \\
\textbf{Comments}   & Notes or comments about the quality or issues identified in the spectral or structural data, such as missing axes or unclear lines. \\
\textbf{EXAFS Type} & Specifies the type of EXAFS data presented, whether it is raw signal data or Fourier transformed data. \\
\textbf{DOI}        & Digital Object Identifier for each article, ensuring traceability back to the source. \\
\textbf{URL}        & Direct URLs to the articles for quick access. \\
\bottomrule
\end{tabular}
\end{table}

Another spreadsheet records the processes involved in extracting iron XAS spectra. It highlights various challenges encountered, such as missing axes on figures and ambiguities in figure interpretation. Such detailed records are crucial for enabling future researchers to understand and evaluate the decision-making processes affecting data inclusion or exclusion. Notably, we include and distinguish between EXAFS data represented in both k-space and r-space. This enhancement ensures that researchers can access and utilize the specific type of data they require for their analyses. We have meticulously labeled each dataset entry in our documentation files to reflect this categorization, addressing a critical need for clarity. The details of this documentation are explained in Table \ref{tab:xas_ann}, and Figure \ref{fig:eg_xas_doc} shows an example of an XAS spectrum data documentation log.

Data entries within our datasets were cross-verified against original publications to ensure accuracy and reliability. Spectral and structural data were, wherever feasible, compared against established databases to ensure consistency. Discrepancies and significant findings are carefully documented. The methodologies employed for data extraction are thoroughly detailed, including the tools used, such as the WebPlotDigitizer for converting plot images to numerical data. Each dataset entry is accompanied by extensive metadata such as detailed URLs of original publications, covering all aspects from data extraction to final dataset compilation. This approach not only enhances reproducibility but also facilitates the integration of our data with other datasets. It ensures a clear understanding of the context and specifics of the data collected.

\newpage

\end{document}